\documentclass[12pt]{iopart}
\newcommand{\no}{\nonumber}
\newcommand{\veps}{\varepsilon}

\usepackage{graphicx}% Include figure files
\usepackage{dcolumn}% Align table columns on decimal point
\usepackage{bm}% bold math

\begin{document}

%\draft

\title{Large-scale electronic structure theory
for simulating nanostructure process}

\author{T. Hoshi$^{1,2}$}
\author{T. Fujiwara$^{1,2}$}
\address{$^{1}$ Department of Applied Physics, University of Tokyo,7-3-1 Hongo, Bunkyo-ku, Tokyo 113-8656, Japan}
\address{$^{2}$ Core Research for Evolutional Science and Technology  (CREST-JST), Japan Science and Technology Agency, 4-1-8 Honcho, Kawaguchi-shi, Saitama 332-0012, Japan}
%\ead{hoshi@coral.t.u-tokyo.ac.jp}
\date{\today}

\begin{abstract}
Fundamental theories and practical methods 
for large-scale electronic structure calculations are given,
in which the computational cost is proportional to the system size. 
Accuracy controlling methods for microscopic freedoms 
are focused
on two practical solver methods,
Krylov-subspace method and
generalized-Wannier-state method.
A general theory called the \lq multi-solver' scheme 
is also formulated,
as a hybrid  between different solver methods.
Practical examples are carried out
in several insulating and metallic systems 
with  $10^3$-$10^5$ atoms. 
All the theories provide 
general guiding principles of 
constructing an optimal calculation 
for simulating nanostructure processes, 
since a nanostructured system 
consists of several competitive regions,
such as bulk and surface regions, 
and the simulation is designed to reproduce the competition
with an optimal computational cost.
\end{abstract}

\maketitle

%%%%%%%%%%%%%%%%%%%%%%%%%%%%%%%%%%%%%%%%%%%%%%%%%%%%%%%%%%%%%%%%%%%%%%%%%
\section{Introduction \label{INTRO}}

\begin{figure}[tbh]
\begin{center}
  \includegraphics[width=7cm]{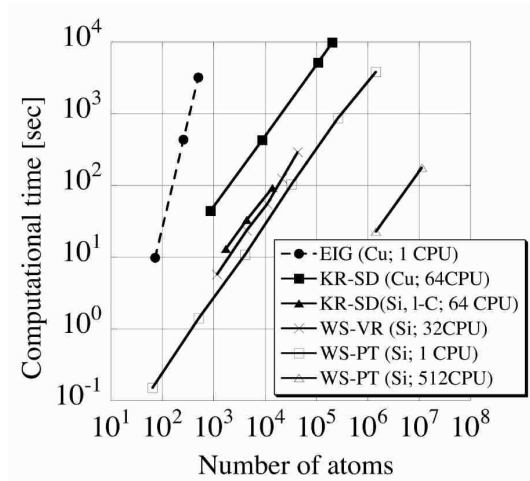}
\end{center}
\caption{\label{FIG-CPU-ORDER-N}
The computational time
as a function of the number of atoms ($N$)
(Refs. \cite{HOSHI2003A, HOSHI2005A}, this work);
Several metallic (fcc Cu and liquid C) and insulating (bulk Si) 
systems are calculated  up to 11,315,021 atoms . 
The time was measured  
for the electronic structure calculation 
with a given atomic structure.
The calculations were carried out
by the conventional eigenstate calculation (EIG) 
and by our methods for large systems; 
(i) Krylov-subspace method 
with subspace-diagonalization procedure (KR-SD), 
(ii) Wannier-state method with variational procedure (WS-VR)
and (iii) Wannier-state method with perturbative procedure (WS-PT). 
For \lq 1CPU' computations, 
we used single Pentium 4$^{\rm TM}$ processor in 2 GHz. 
Parallel computations were carried out 
by SGI Origin 3800$^{\rm TM}$ (for WS-PT method), 
Origin 2800$^{\rm TM}$ (for WS-VR method) and
Altix 3700$^{\rm TM}$  (for KR-SD method). 
See text for details.
%The parallelism is done by the OpenMP technique 
%(www.openmp.org). 
}
\end{figure}

Electronic structure theory 
plays a crucial role in understanding and controlling 
nanostructures, 
structures in nano-meter and ten-nano-meter scales.
Dynamical simulation in these scales is, however, 
impractical for the present standard methodology,
such as the Car-Parrinello method \cite{CP},
owing to its heavy computational cost.
From 1990's,
many calculation methods  and related techniques
have been proposed  
for large systems, systems with thousands of atoms or more, 
by calculating one-body density matrix or the Green's function,
instead of one-electron eigenstates.
\cite{KOHN96, REVIEW-ON, REVIEW-ON2, MAURI, LNV1993, ORDEJON1993, 
Goedecker1994, HOSHI1997, 
Roche-Mayou, HOSHI2000A, Ozaki-Terakura2001, SIESTA, Bowler, 
HOSHI2003A, TAKAYAMA2004A, HOSHI2005A, ONETEP, TAKAYAMA2006}
In these methodologies, 
calculation is carried out with real-space representation and 
a physical quantity $\langle X \rangle $ is given as a trace form
\begin{equation}
\langle X \rangle 
 = {\rm Tr}[ \rho X ]
  = \int\int d{\bm r}d{\bm r}^\prime 
  \rho ({\bm r},{\bm r}^\prime ) X({\bm r}^\prime, {\bm r}).
\label{TRACE-EQ}
\end{equation}
Here the one-body density matrix $\rho$ is defined, 
from occupied one-electron eigenstates ${\phi_k(\bm{r})}$, as 
\begin{equation}
\ \rho \equiv \sum_k^{\rm occ.} | \phi_k^{\rm (eig)} 
\rangle \langle \phi_k^{\rm (eig)}  | .
\label{DM-DEF}
\end{equation}
One can find that, if the matrix $X({\bm r},{\bm r}^\prime)$ is of short range,
the off-diagonal long-range component of the density matrix 
does not contribute to the physical quantity $\langle X \rangle $,
which is important for practical success of large-scale calculations. ~\cite{KOHN96}
Actual calculation methods and their applications are found in 
recent reviews \cite{REVIEW-ON, REVIEW-ON2} 
or papers.~\cite{MAURI, LNV1993, ORDEJON1993, 
Goedecker1994, HOSHI1997, 
Roche-Mayou, HOSHI2000A, Ozaki-Terakura2001, SIESTA, Bowler, 
HOSHI2003A, TAKAYAMA2004A, HOSHI2005A, ONETEP, TAKAYAMA2006}
A set of theories and program codes have been developed in our group 
and a test calculation of Fig.~\ref{FIG-CPU-ORDER-N} shows 
that the computational cost is \lq order-$N$' or proportional 
to the system size ($N$) among the calculations 
with $10^3$-$10^7$ atoms
\cite{HOSHI2003A, HOSHI2005A, TAKAYAMA2004A, HOSHI2005B,TAKAYAMA2006}.

A practical success in an application study always requires 
the balance between the accuracy and the computational cost. 
Every calculation method has several controlling parameters 
and one should establish a systematic way 
of setting them in optimal values. 
Here we remember that  
a nanostructure is 
composed of several comparable regions 
with essential difference in electronic structure,
such as  bulk and surface regions. 
Since the competition of these regions
gives various structural and functional properties of nanostructures,
the requirement on dynamical simulation of a nanostructure 
is to reproduce the competition,
or to reproduce the difference in electronic structure among the regions,
throughout the process. 

In this paper, we will show  
how to construct an optimal calculation scheme 
for nanostructure process. 
The essential concepts are 
(i) controlling method of the accuracy and the computational cost
by monitoring residuals for  microscopic or basis freedoms 
and 
(ii) choice or combination of different calculation methods.
Hereafter 
the word \lq solver method' is used 
as a practical calculation method 
of density matrix $\rho$ with a given Hamiltonian $H$.

This paper is organized as follows;
In Sec.~\ref{THEORY-SOLVER}, we will explain the foundation of two methods,
Krylov subspace method 
and generalized Wannier state method.
They are practical solver methods to calculate 
the density matrix for a given system
and we will compare them, in Sec.~\ref{SEC-COMPARI},
from a practical view point.
In Sec.~\ref{SEC-HYB}, 
we will construct a methodology 
of \lq multi-solver' scheme,
as a hybrid or combination of different solver methods.
Several applications as molecular dynamics (MD) simulations
will be presented  in Sec.~\ref{SEC-APPL},
so as to clarify the methodological points. 
In the present paper,
we limit the formulations 
into those for a Hamiltonian $H$ as a real-symmetric matrix.
Practical calculations 
were carried out with Hamiltonians 
in the Slater-Koster (tight-binding) form;
The Hamiltonian for fcc Cu is constructed 
from the first-order form $H^{(1)}$ of 
the linear muffin-tin orbital theory~\cite{LMTO}
and those for C and Si are typical ones in
Ref.~\cite{XU} and Ref.~\cite{KWON}, respectively.

%............(not yet)................

%-%-%-%-%-%-%-%-%-%-%-%-%-%-%-%-%-%-%-%-%-%-%-%-%-%-%-%-%-%-%
%-%-%-%-%-%-%-%-%-%-%-%-%-%-%-%-%-%-%-%-%-%-%-%-%-%-%-%-%-%-%
%%%%%%%%%%%%%%%%%%%%%%%%%%%%%%%%%%%%%%%%%%%%%%%%%%%%%%%%%%%%%%%%%%%%%%%%%
\section{Theory (1) Practical solver methods}
\label{THEORY-SOLVER}

%%%%%%%%%%%%%%%%%%%%%%%%%%%%%%%%%%%%%%%%%%%%%%%%%%%%%%%%%%%%%%%%%%%%%%%%%
\subsection{Solver methods with Krylov subspace}
\label{KRYLOV}

Krylov subspace is
a general mathematical concept  
defined as the linear space of 
\begin{equation}
\label{eq:krylov1}
{\cal K}_{\nu}(H, | j \rangle) \equiv {\rm span}\left\{
|j \rangle, \  H| j \rangle, \  H^2 |j \rangle, \  \ldots, \  H^{\nu_{}-1}|j  \rangle
\right\}.
\end{equation}
Here the \lq starting' vector ($|j \rangle$) and 
the dimension of the subspace ($\nu$) are arbitrary. 
Many iterative methods, 
such as the standard conjugate-gradient method,
are formulated with Krylov subspace.
See a recent textbook~\cite{KRYLOV}, for example.
In the present context, the matrix $H$ is a Hamiltonian
and $|j \rangle$ is a real space basis.
A large-scale calculation can be realized,
when the  density matrix 
$\langle i | \rho | j \rangle $ 
is constructed 
within the Krylov subspace ${\cal K}_{\nu}(H, | j \rangle)$.
The Krylov subspace method enables us also to calculate
the Green's function
$\langle i | G | j \rangle $, 
which gives directly the information of electronic states, 
such as the density of states (DOS).  
%which is called \lq subspace diagonalization'.
When the dimension $\nu$ is equal to that of the original Hamiltonian matrix $H$,
the linear space of Eq.~(\ref{eq:krylov1}) is complete and  
all the calculation results are exact. \cite{KRYLOV}

\subsubsection{Subspace-diagonalization method}
\label{KRYLOV-SD} 

Here we explain 
a practical solver method with Krylov subspace,  
called \lq subspace-diagonalization method' (KR-SD) \cite{TAKAYAMA2004A};
First, we construct an orthogonal basis set $\{ | K_n^{(j)} \rangle \}$ 
for the Krylov subspace ($\langle K_{n}^{(j)}  | K_{m}^{(j)}\rangle =\delta_{nm}$);
\begin{eqnarray}
 {\cal K}_{\nu}(H, | j \rangle)  =  
{\rm span} \, \{   | K_1^{(j)} \rangle \equiv | j \rangle , \, \, | K_2^{(j)} \rangle , \cdot \cdot , 
| K_{\nu}^{(j)}\rangle  \}  
\end{eqnarray}
by the Lanczos procedure, a three-term recurrence formula. 
The $n$-th basis $ | K_n^{(j)} \rangle$ is constructed 
in the $n$-dimensional Krylov subspace 
($| K_n^{(j)} \rangle \in {\cal K}_{n}(H, | j \rangle)$). 
In result, a reduced Hamiltonian matrix 
\begin{eqnarray}
\left( H^{{\rm K}(j)} \right)_{nm} \equiv    
\langle K_n^{(j)}  |  H^{(j)}  | K_m^{(j)} \rangle
\end{eqnarray}
is obtained as an explicit $( \nu \times \nu )$ matrix.  
A typical subspace dimension is $\nu=30$ in MD simulations.
Then, we diagonalize the reduced (small) matrix
\begin{eqnarray}
H^{{\rm K}(j)} | v_{\alpha}^{(j)} \rangle = \varepsilon_{\alpha}^{(j)}  | v_{\alpha}^{(j)} \rangle, 
\label{LABEL-KR-EIG}
\end{eqnarray}
with a negligible computational cost. 
The resultant eigen vectors $| v_{\alpha}^{(j)} \rangle$ are described as the set of coefficients 
$C_{\alpha m}^{(j)} \equiv \langle K_m^{(j)}  | v_{\alpha}^{(j)} \rangle$. 

The density matrix is obtained by
\begin{eqnarray}
\langle i | \rho | j \rangle &\Rightarrow &
\langle i |  \rho^{K(j)} | j \rangle \\
 \label{KR-DEF-DM2}
&=& \sum_n^{\nu} 
\langle i |  K_{n}^{(j)} \rangle \langle K_{n}^{(j)} |{\rho}_{}^{K(j)} |  j \rangle, 
\label{EQ-DM-SUMMATION}
\end{eqnarray}
with the definition of 
\begin{eqnarray}
 \rho^{K(j)} \equiv \sum_{\alpha}  | v_{\alpha}^{(j)} \rangle 
 f_{\tau}(\varepsilon_{\alpha}^{(j)}-\mu)
 \langle  v_{\alpha}^{(j)} |
 \label{KR-DEF-DM1}.
\end{eqnarray}
Here the occupation number 
$f_{\tau}(\varepsilon-\mu )$ is given by
the Fermi-Dirac function with 
a temperature (level-broadening) parameter $\tau$
and the chemical potential $\mu$. 
The chemical potential is determined by the bisection method. 
The Green's function $\langle i | G (z) | j \rangle$ can be calculated 
in a similar manner. \cite{TAKAYAMA2004A} 
%and  is used 
%to obtain the electronic property, such as density of states (DOS).
In short,  
the present procedure is a standard quantum-mechanical calculation 
for eigen states, except the point that 
the calculation is carried out within the Krylov subspace.
Therefore it is straightforward 
to apply the method to calculations with a nonorthogonal basis set,  
in which 
a generalized eigen-value equation,
instead of Eq.~(\ref{LABEL-KR-EIG}), 
is solved within the subspace.

\subsubsection{Shifted conjugate-orthogonal conjugate-gradient method}
\label{KRYLOV-SCOCG} 

Another solver method with Krylov subspace
was formulated and called 
\lq shifted conjugate-orthogonal conjugate-gradient' (SCOCG) method.
\cite{TAKAYAMA2006}
Its foundation is given by 
a mathematical theorem 
proved recently.~\cite{FROMMER} 
The practical procedure is based on an iterative solver method 
for the linear equation of the  Green's function;
\begin{eqnarray}
 (z-H)   | x_j \rangle = | j \rangle,
\end{eqnarray}
because of 
$G_{ij} =  \langle i  | x_j \rangle = \langle i | (z-H)^{-1}  | j \rangle$. 
The density matrix is obtained by 
\begin{equation}
 \rho_{ij} = -\frac{1}{\pi} \int_{-\infty}^{\infty}
 {\rm Im} \, G_{ij}(\varepsilon + i 0) \, 
f_{\tau}(\varepsilon-\mu ) \, {\rm d}\varepsilon. 
 \label{eq:rho_int}
\end{equation}
The SCOCG method and KR-SD method
share many common features
but are different in the numerical treatment. 
See the original paper \cite{TAKAYAMA2006} for detailed comparison.
For the present time, 
we use, mainly, the KR-SD method for MD simulation
and we think that 
the SCOCG method is suitable to discuss a very fine energy spectrum 
of the Green's function. \cite{TAKAYAMA2006}

\subsubsection{Accuracy control with residual}
\label{KRYLOV-ACCURACY} 

So as to monitor the  accuracy during the simulation,
we calculate 
a residual vector
of the Green's function; \cite{TAKAYAMA2006}
\begin{eqnarray}
|  \delta G_j  \rangle \equiv (z-H) G  | j \rangle -  | j \rangle.
\label{EQ-RN}
\end{eqnarray}
The residual vector is defined
individually among the basis suffix $j$.
We observed that 
the required subspace dimension  $\nu=\nu^{(j)}$ 
for a given criteria on the residual vector  
is different among surface and bulk regions. \cite{TAKAYAMA2006}
Such a determination of the controlling parameters $\{ \nu^{(j)} \}$
is an example of the accuracy control for microscopic or basis freedoms.

%%%%%%%%%%%%%%%%%%%%%%%%%%%%%%%%%%%%%%%%%%%%%%%%%%%%%%%%%%%%%%%%%%%%%%%%%
\subsection{Solver methods with generalized Wannier state \label{SEC-WANI-ON}}
%\subsection{Theory}

Another method for obtaining the density matrix in large systems
is formulated using generalized Wannier state.
\cite{KOHN-WANI73,KOHN-WANI93,
MAURI, ORDEJON1993, MARZARI, HOSHI2000A, HOSHI2001A, ANDERSEN2003}
%a generalization of the (conventional) Wannier state. \cite{WANI, WANI2}
A physical picture of the generalized Wannier states 
is localized chemical wave function in condensed matters,
such as a bonding orbital or a lone-pair orbital
with a slight spatial extension or \lq tail'.
\cite{MAURI, ORDEJON1993, MARZARI, HOSHI2000A, HOSHI2001A, ANDERSEN2003}
Their wavefunctions $\{  \phi_i^{\rm (WS)}  \}$
are equivalent to the unitary transformation of occupied eigen states
and satisfy the equation of 
\begin{eqnarray}
 H | \phi_i^{\rm (WS)} \rangle
 = \sum_{j=1}^{\rm occ} \varepsilon_{ij} | \phi_j^{\rm (WS)} \rangle, 
 \label{REV-SCE-UNITARY}
\end{eqnarray}
where the matrix $\varepsilon_{ij}$ is 
the Lagrange multipliers 
for the orthogonality constraint 
$(\langle \phi_i^{\rm (WS)}  | \phi_j^{\rm (WS)} \rangle = \delta_{ij})$.
The suffix $i$ of a wavefunction $\phi_i^{\rm (WS)}$ 
indicates the position of its localization center, such as bond site.
Since Wannier states
give the density matrix $\rho$ in Eq.~(\ref{DM-DEF}) 
by replacing eigen states 
$\{ \phi_k^{\rm (eig)}\}$ into  Wannier states 
$\{ \phi_i^{\rm (WS)}\}$, 
any physical quantity can be reproduced
in the trace form of Eq.~(\ref{TRACE-EQ}). 

Our practical solver methods
are based on a mapped eigen-value equation 
\cite{HOSHI2000A,THESIS} 
that is  equivalent to 
Eq.~(\ref{REV-SCE-UNITARY}); 
\begin{eqnarray}
 H_{\rm WS}^{(i)}  | \phi _i^{\rm (WS)} \rangle 
 = \veps_{\rm WS}^{(i)}  | \phi _i^{\rm (WS)} \rangle,
 \label{MFE}
\end{eqnarray}
where 
\begin{eqnarray}
& &  H_{\rm WS}^{\rm (i)} \equiv 
 H  + 2 \eta_{\rm s} \bar{\rho}_i - H  \bar{\rho}_i - \bar{\rho}_i H  
 	 \label{EQ-WANI-HAMI} \\
& & \bar{\rho}_i 
     \equiv  \rho -  | \phi_i^{\rm (WS)} \rangle \langle \phi_i^{\rm (WS)} | 
	 =\sum_{j (\ne i)}^{\rm occ.} | \phi_j^{\rm (WS)} \rangle \langle \phi_j^{\rm (WS)} |. 
\end{eqnarray}
The energy parameter $\eta_{\rm s}$ should be much larger
than the highest occupied level. 
%The proof is given in Refs.~\cite{HOSHI2000A} and Appendix XX
%with different ways of proof.
Equation~(\ref{MFE}) was derived in Refs.~\cite{HOSHI2000A,THESIS}
and will be derived again, from a different theoretical background, 
in Sec.~\ref{SEC-HYB-THEORY} of this paper. 

\subsubsection{Variational procedure in Wannier-state method}
\label{WS-VR}

Equation (\ref{MFE}) gives
a practical iterative procedure
to generate Wannier states
under explicit localized constraint, \cite{HOSHI2000A, HOSHI2003A,HOSHI2005A,THESIS}
which is called
variational Wannier state method.
See papers \cite{HOSHI2000A,THESIS} for details.
Residual vector for each wavefunction $| \phi_i^{\rm (WS)} \rangle$
\begin{eqnarray}
| \delta \phi_i^{\rm (WS)} \rangle \equiv 
 H_{\rm WS}^{(i)}  | \phi _i^{\rm (WS)} \rangle - \veps_{\rm WS}^{(i)}  | \phi _i^{\rm (WS)} \rangle.
 \label{WANI-RV}
\end{eqnarray}
is monitored  for each Wannier state 
during the simulation, 
so as to control the accuracy,
which realizes 
the accuracy control for microscopic or basis freedoms, 
as discussed  in the Krylov-subspace method with Eq.~(\ref{EQ-RN}).
A practical success in the Wannier-state method is realized, 
when all or a dominant number of wavefunctions are well localized. 
Examples and technical details are given in 
Sec.~\ref{APPL-SILICON-FRAC} and references 
\cite{HOSHI2000A, HOSHI2003A, THESIS}.

\subsubsection{Perturbative procedure in Wannier-state method}
\label{WS-PT}

We developed also a perturbative method,
~\cite{HOSHI2000A, HOSHI2001A, GESHI, THESIS}
in which a perturbation solution
of  Eq.~(\ref{MFE}) is constructed 
for each Wannier state $| \phi_i^{\rm (WS)} \rangle$;
\begin{eqnarray}
| \phi_i^{\rm (WS)} \rangle \Rightarrow 
C_i \left( | \phi_i^{\rm (WS)(0)} \rangle +  
| \phi_i^{\rm (WS)(1)} \rangle\right).
 \label{WANI-PTWFN}
\end{eqnarray}
Here $| \phi_i^{\rm (WS)(0)} \rangle$
and $| \phi_i^{\rm (WS)(1)} \rangle$ are 
the unperturbed and first-order perturbation terms, respectively,
and the factor $C_i$ is the normalization factor. 
The unperturbed term $| \phi_i^{\rm (WS)(0)} \rangle$
should be prepared as an input quantity 
and the perturbation term $| \phi_i^{\rm (WS)(1)} \rangle$
and the normalization 
factor $C_i$
are determined automatically 
by the standard first-order perturbation procedure. 
~\cite{HOSHI2000A, HOSHI2001A, GESHI, THESIS} 
During a simulation, 
the weight of the unperturbed term 
\begin{eqnarray}
w_{0}^{(i)} \equiv 
| \langle  \phi_i^{\rm (WS)(0)} | \phi_i^{\rm (WS)} \rangle|^2 
 \label{WANI-PT-WT}
\end{eqnarray}
is monitored, 
for each wavefunction, 
as an accuracy control
for microscopic or basis freedoms.
In silicon crystal, for example,
the ideally $sp^3$-bonding wavefunction 
is chosen as the unperturbed term
and the weight of the  unperturbed term 
is dominant ($w_{0}^{(i)} =0.94$)
~\cite{HOSHI2000A, HOSHI2001A, THESIS},
which validates the perturbative treatment.
When the perturbative method is validated, 
its computational performance 
is faster than that of the variational method, 
since the perturbative method 
gives a simpler procedure to generate the wavefunctions 
and does not require any iteration loop.

\subsection{Comparison between Krylov-subspace and Wannier-state methods}
\label{SEC-COMPARI}

When the Wannier-state methods are compared with the Krylov-subspace methods,
the Wannier-state methods require
an initial guess of wavefunctions
in the variational (iterative) method or 
an unperturbed term of wavefunction 
in the perturbative method. 
As an example,
the reconstruction on Si(001) surface was calculated 
with the force on atoms.
The calculation was carried out 
by the two Krylov-subspace methods,
(i) the subspace diagonalization procedure \cite{TAKAYAMA2004A}  and 
(ii) the  SCOCG procedure \cite{TAKAYAMA2006}, 
and (iii) the variational Wannier-state method. \cite{THESIS} 
In the variational Wannier-state method, 
the initial guess of the wavefunctions are prepared to be 
the lone-pair state of $(| s \rangle + | p_z \rangle )/\sqrt{2}$
for surface states 
and to be the $sp^3$-bonding states for other (bulk) states.
The three methods
reproduce  the energy differences satisfactorily 
among 
the $(2 \times  1)$, $(2 \times  2)$, and $(4 \times  2)$ surfaces, 
when these results are compared with
those of  
the eigen-state calculation with the present Hamiltonian \cite{FU} 
and the standard {\it ab initio} calculation \cite{RAMSTAD}.
%while such a preparation procedure of input quantity is 
%not needed for the Krylov-subspace method.

The perturbative Wannier-state method 
is much limited in its applicability than the above three methods,
because 
the unperturbed term  
should be prepared as a good approximation
($| \phi_i^{\rm (WS)} \rangle \approx
| \phi_i^{\rm (WS)(0)} \rangle $ 
or $w_{0}^{(i)} \approx 1$).  
So far we have applied the perturbative Wannier-state method
only to the bulk ($sp^3$-bonding) wavefunction 
in the diamond-structure solids
without deformation 
or with small (elastic) deformation. 
~\cite{HOSHI2000A, HOSHI2001A, GESHI, THESIS} 
Since the first-order perturbation form was used
for the wavefunction $ | \phi_i^{\rm (WS)} \rangle$  in these cases, 
the calculated energy $\varepsilon_i^{\rm (WS)} 
\equiv \langle  \phi_i^{\rm (WS)} | H | \phi_i^{\rm (WS)} \rangle$
is correct within the second order 
with respect to deformation and  
the elastic constants are well reproduced.
We should say, however, that 
a drastic change of wavefunction, 
like that in a bond-breaking process,  
is not reproduced by the perturbative Wannier-state method,
if the bulk ($sp^3$-bonding) wavefunction is chosen 
as the unperturbed term. 

Despite the limitations, 
the computational performance of the Wannier-state methods 
is faster, at best by several hundred times, 
than that of the Krylov-subspace method, 
if it is applicable. 
In Fig.~\ref{FIG-CPU-ORDER-N}, for example, 
the Wannier-state method with 
the perturbative procedure 
(WS-PT) using single CPU is faster than 
the Krylov-subspace method with 
the subspace-diagonalization procedure 
(KR-SD) using 64 CPUs.  

When one thinks about a guiding principle 
for how to choose a solver method in an application study,
the above discussion suggests
that the Wannier-state methods 
give a faster performance, 
when the input wavefunctions are near the final solutions
and, particularly,  they are well localized. 
In other cases,  
the Krylov-subspace method is preferable,
since the Krylov-subspace method  does not require 
any input quantity for electronic states.

%an optimal method 

%on a guiding principle for choice of the solver method.

\section{Theory (2) Multi-solver scheme}
%%%%%%%%%%%%%%%%%%%%%%%%%%%%%%%%%%%%%%%%%%%%%%%%%%%%%%%%%%%%%%%%%%%%%%%%%
%\subsection{Multi-solver scheme}
\label{SEC-HYB}
 
\subsection{Formulation}
\label{SEC-HYB-THEORY}
 
 As another fundamental methodology for large-scale calculations,
we developed a \lq multi-solver' scheme  
\cite{HOSHI2003A,THESIS},
as hybrid or combination of two different solver methods.
Its basis idea is that 
the density matrix is decomposed into two parts
called \lq subsystems' and 
they are given by  different solver methods.
As discussed below,  
the multi-solver scheme will be fruitful, particularly,
when the simulation system is composed 
of different regions, such as bulk and surface regions,
and different solver methods are used in these different regions.

The mathematical foundation of the multi-solver scheme
is based on the commutation relation of the density matrix;
\begin{eqnarray}
[ H,  \rho]=0.
 \label{HYB-EQN-0}
\end{eqnarray}
When the occupied one-electron states, 
eigenstates or Wannier states, 
are classified into two groups A and B,
the density matrix is decomposed into the corresponding two parts
\begin{eqnarray}
\rho \equiv  \rho_{\rm A} + \rho_{\rm B}
\end{eqnarray}
where 
\begin{eqnarray}
& & \rho_{\rm A} \equiv  
\sum_i^{\rm occ. (A)}  | \phi _i  \rangle    \langle \phi _i |, 
    \label{HYB-RHOA-EIG-DEV}  \\
& & \rho_{\rm B} \equiv  
\sum_j^{\rm occ. (B)}  | \phi _j  \rangle    \langle \phi _j |  
 = \rho - \rho_{\rm A}.
 \label{HYB-RHOB-EIG-DEV}
\end{eqnarray}
Here we call $\rho_A$ and $\rho_B$ \lq subsystems' 
and the two subsystems are orthogonal
\begin{eqnarray}
 \rho_{\rm A}  \rho_{\rm B}  = 0,
 \label{HYB-ORTHO}
\end{eqnarray}
owing to the orthogonality relation
\begin{eqnarray}
\langle \phi _i | \phi _j \rangle = 0, \quad
\phi_i \in {\rm A}, \phi_j \in {\rm B}.
 \label{HYB-WFN-ORTHO}
\end{eqnarray}

If the subsystems $\rho_A, \rho_B$ are defined from eigenstates, 
a mapped Hamiltonian 
\begin{eqnarray}
 H_{\rm map}^{\rm (A)} \equiv H + 2 \eta_{\rm s} \rho_{\rm B} 
 \label{HYB-HAM-A},
\end{eqnarray}
with a scalar $\eta_{\rm s}$,
satisfies  the commutation relation
\begin{eqnarray}
[ H_{\rm map}^{\rm (A)},  \rho_{\rm A}] = 0
 \label{HYB-EQN-A}
\end{eqnarray}
owing to Eq.~(\ref{HYB-ORTHO}) and 
\begin{eqnarray}
& &  [ H,  \rho_{\rm \alpha}] = 0 \quad (\alpha=A,B).
  \label{HYB-EQN-AB1} 
\end{eqnarray}
We call the scalar $\eta_{\rm s}$  \lq energy-shift parameter'. 
If $\rho_B$ is given,
the problem for obtaining $\rho_{\rm A}$
is reduced to 
a standard quantum mechanical problem 
with the well-defined Hamiltonian 
$H_{\rm map}^{\rm (A)}$.
%and the electron number 
%$N_{\rm A} \equiv {\rm Tr}[\rho_{\rm A}]$.
In practical calculations, 
the energy shift parameter is chosen to be 
so large 
that the states in $\rho_B$ do not lie
%and only the states in $\rho_A$ appear 
in the occupied energy region of $H_{\rm map}^{\rm (A)}$.
Note that Eqs.~(\ref{HYB-EQN-A}), (\ref{HYB-EQN-AB1})  are satisfied, 
even if the subsystems $\rho_A, \rho_B$ are constructed
by eigen states with fractional occupancy. 

If the subsystems $\rho_A, \rho_B$ are defined from 
Wannier states, on the other hand, 
Eq.~(\ref{HYB-EQN-A}) are not satisfied,
because Eq.~(\ref{HYB-EQN-AB1}) is not satisfied.  
Then we redefine the mapped Hamiltonian as
\begin{eqnarray}
 H_{\rm map}^{\rm (A)} \equiv H + 2 \eta_{\rm s} \rho_{\rm B} 
 -  H \rho_{\rm B} -  \rho_{\rm B}  H,
 \label{HYB-HAM-1-ALL}
\end{eqnarray}
which satisfies Eq.~(\ref{HYB-EQN-A}) 
in the cases of eigenstates and Wannier states.
See  \ref{APPENDIX-HYB-WANI} for proof. 
A simplest case is that 
the subsystem $\rho_A$ is consist of only one  Wannier state $\phi_i^{\rm (WS)}$.
%and all the other  Wannier states are contributed to the subsystem B.
In this case,
the mapped Hamiltonian in Eq.~(\ref{HYB-HAM-1-ALL}) 
is reduced to that in of Eq.~(\ref{EQ-WANI-HAMI}),
because of  
$\rho_{\rm A} \Rightarrow | \phi_i^{\rm (WS)} \rangle \langle \phi_i^{\rm (WS)} |$ 
and
$\rho_{\rm B} \Rightarrow \bar{\rho}_i$.
In other words, 
the present theory gives 
another derivation of Eq.~(\ref{MFE}), 
the mapped equation of Wannier state.
From a practical view point, 
the term $(H \rho_{\rm B} +  \rho_{\rm B}  H)$
in Eq.~(\ref{HYB-HAM-1-ALL}) can be ignored,
when the energy shift parameter $\eta_{\rm s}$ is so large
$(\eta_{\rm s} \rightarrow +\infty)$
that the energy band of $\rho_B$ is well separated, energetically,
from that of $\rho_A$. 
If the term is ignored, 
the mapped Hamiltonian is reduced to 
the form of Eq.~(\ref{HYB-HAM-A}).

%%%%%%%%%%%%%%%%%%%%%%%%%%%%%%%%%%%%%%%%%
\begin{figure}[ht]
\begin{center}
 \includegraphics[width=14cm]{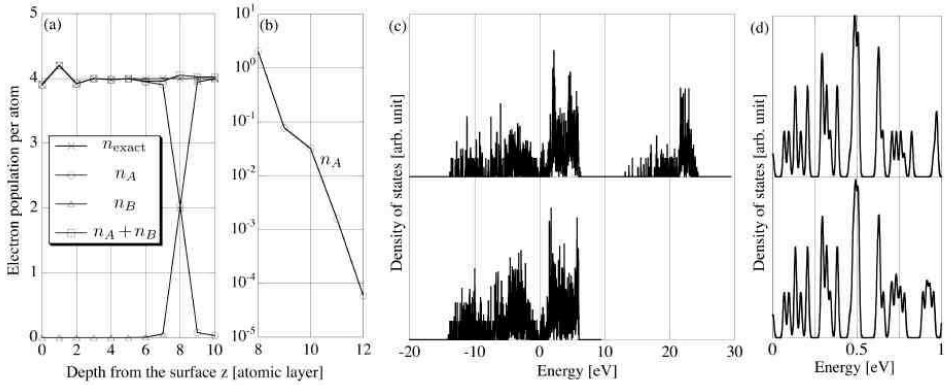}
\end{center}
\caption{
Example of the multi-solver scheme 
in a silicon slab with ideal (001) surface;
(a)(b)The electron population per atom is plotted 
as the function of atomic layer.
The atoms at $z=0$ correspond to the surface atoms.
The calculation is carried out by 
the conventional eigen-state calculation ($n_{\rm exact}$)
and the present multi-solver scheme ($n_A$ and $n_B$).
(c)(d) DOS in the multi-solver scheme.
Lower panel : DOS of the original Hamiltonian $H$.
Upper panel : DOS of the mapped Hamiltonian $H_{\rm map}^{\rm (A)}$.
%The system is a silicon slab with (001) surface.
}
\label{fig-hybrid-tot-density}
\end{figure}
%%%%%%%%%%%%%%%%%%%%%%%%%%%%%%%%%%%%%%%%%

%%%%%%%%%%%%%%%%%%%%%%%%%%%%%%%%%%%%%%%%%
\subsection{Example 1}
\label{HYB-EX1}

Hereafter, the multi-solver scheme will be demonstrated. 
Although the formulation of the multi-solver scheme is general,
we have used, so far, the scheme 
only with the perturbative Wannier-state method for 
a subsystem ($\rho_B$). 
Among these cases, 
the subsystem $\rho_B$ is determined
in the first-order perturbation form, and then 
the other subsystem $\rho_A$ is determined, 
through the mapped Hamiltonian 
$H_{\rm map}^{(A)}$,
by a different solver method  
($\rho_B \Rightarrow H_{\rm map}^{(A)} \Rightarrow \rho_A$).
In other words, the present procedure does not contain
a self-consistent loop 
($\rho_B \Rightarrow H_{\rm map}^{(A)} \Rightarrow \rho_A
\Rightarrow H_{\rm map}^{(B)}\Rightarrow \rho_B \cdot \cdot$).
A related general discussion will be given in 
Sec.~\ref{HYB-DISCUSS}. 

The first example is  a Si slab with ideal (001) surface, 
in which we use the eigen-state method for $\rho_A$
and the perturbative Wannier-state method for $\rho_B$.  
Each atomic layer contains 64 atoms and 
the total number of atoms is $64 \times 16 = 1024$ 
in the periodic simulation cell.
Since an ideal (001) surface gives
an almost zero energy gap (0.025 eV), 
the present example is one of the severest tests
for the present methodology. 
The $z$ coordinate is written 
in the unit of atomic layer ($z=0,1,2....15$).
The surface atoms are located at $z=0$ and 
have dangling-bond electrons.
The atoms in the opposite surface  ($z = 15$) 
are terminated 
by the bulk (sp$^3$-bonded) Wannier states
and do not have any dangling-bond electrons.
The $z$ coordinate of bulk-bond sites can be described 
as half integers $(z=0.5, 1.5, 2.5, ....14.5)$. 
In the multi-solver scheme, 
the subsystem $\rho_B$ is constructed 
from the Wannier states whose 
localization  (bond) centers 
are located deeper than 
the eighth atomic layer ($z = 8.5, 9.5, ..$).
The rest system is assigned 
to the subsystem $\rho_A$ that 
contains the surface states. 
The wavefunctions $\phi _i^{({\rm WS})}$  in $\rho_B$
are determined 
by the perturbation form and,
then, $\rho_A$ is determined 
by diagonalizing the mapped Hamiltonian $H_{\rm map}^{(A)}$.
The energy shift parameter is chosen as 
$2 \eta_{\rm s} = 1$a.u. ($\approx$ 27.2 eV).

In Fig.~\ref{fig-hybrid-tot-density}(a),
%the electron population per atom is 
%plotted as the function of the atomic coordinate $z$.
the electron populations of the subsystems, $n_A(z)$ and $n_B(z)$, 
are plotted as the function of the atomic coordinate $z$.
%and compared to that of the exact calculation $n_{\rm exact}(z)$.
The total electron population in the multi-solver scheme
$(n_A + n_B)$ reproduces  the exact one $n_{\rm exact}(z)$. 
%which corresponds to the \lq tail' part 
%of the Wannier states in the subsystem $\rho_A$.
As a remarkable result, 
the population at $z=8$ 
is contributed by 
both of the subsystem $\rho_A$ and $\rho_B$ 
with an almost equal weight, 
since 
the Wannier states located at $z=7.5$ 
and those at $z=8.5$ belong to
$\rho_A$ and $\rho_B$, respectively. 
Figure \ref{fig-hybrid-tot-density}(b) shows that 
$n_A(z)$ decays quickly at $z > 8$,
because of the nature of 
the mapped Hamiltonian $H_{\rm map}^{(A)}$.

So as to understand the multi-solver scheme,
Figs.~\ref{fig-hybrid-tot-density}(c)(d) show
the DOS of the original Hamiltonian and 
the mapped Hamiltonian $H_{\rm map}^{\rm (A)}$.
The energy origin ($\veps = 0$)
is chosen at the lowest unoccupied level in $H$.
Each eigen level is drawn
as a spike with the width of $\Delta \varepsilon = 0.02$ eV.
In the DOS of the mapped Hamiltonian, 
the band in the occupied energy region
($\veps < 0$) 
is that of $\rho_{\rm A}$, 
while the band of  $\rho_B$ is shifted
by $2 \eta_{\rm s} = 27.2 {\rm eV}$, 
owing to the term of 
$2 \eta_{\rm s} \rho_B$ in $H_{\rm map}^{\rm (A)}$,
and is located at the high-energy region 
at $13 {\rm eV} < \veps < 25 {\rm eV}$.
As in Fig.~\ref{fig-hybrid-tot-density}(d), 
the two DOS profiles agree excellently 
at the bottom of the unoccupied energy region
($0 \le \varepsilon \le 0.7$eV).
%which is reflected from the essence of the multi-solver scheme, 
Here we recall that 
the original and mapped Hamiltonians,
from their definitions,
share the unoccupied eigen states 
that gives the density matrix of 
$\bar{\rho} \equiv 1 - \rho$
($[\bar{\rho}, H]=[\bar{\rho}, H_{\rm map}^{\rm (A)}]=0$)
and the disagreement in the present result
appears only because $\rho_B$ is deviated from the exact one. 
The excellent agreement 
at the band bottom 
($0 \le \varepsilon \le 0.7$eV) appears, 
since these states are contributed dominantly by surface states
and are almost free from the subsystem $\rho_B$.

%%%%%%%%%%%%%%%%%%%%%%%%%%%%%%%%%%%%%%%%%
\begin{figure}[hbt]
\begin{center}
  \includegraphics[width=10cm]{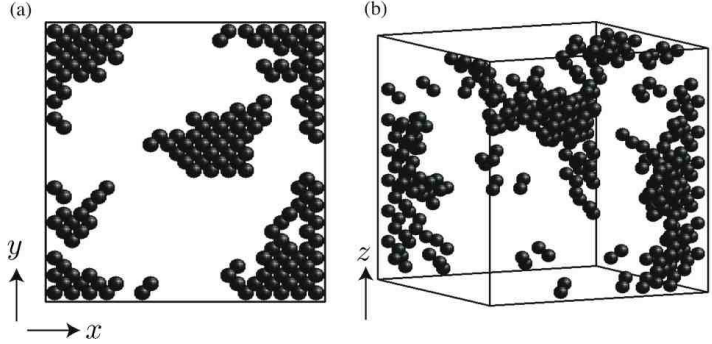}
\end{center}
\caption{
Multi-solver scheme 
with automatic assignment of subsystems ($\rho_A$, $\rho_B$). 
Top (a) and three-dimensional (b) views 
of a silicon nano-crystal with 4501 atoms.
Atoms are visible, only if 
its electron population is dominated by the subsystem $\rho_A$.
The figures are drawn in ideal crystalline geometry for eye guide,
though the actual system is deformed.
The sample edges are plotted as lines for eye guide. 
\label{fig-4501-0716a-hyb-demo}
}
\end{figure}
%%%%%%%%%%%%%%%%%%%%%%%%%%%%%%%%%%%%%%%%%

\subsection{Example 2}

The multi-solver scheme was demonstrated 
in large-scale calculations.
The first example is  reconstructed (001) surface 
of  Si slab with $10^4$ atoms, 
which is determined with the force on atoms.
The practical procedure is the same as in Sec.\ref{HYB-EX1},
except the point that  
the subsystem $\rho_A$ is calculated 
by the Krylov-subspace (KR-SD) method
instead of the exact diagonalization method.
The result shows the correct surface reconstruction. 
\cite{TAKAYAMA2004A}

The second example is  
a MD simulation of 
a silicon nanocrystal with 4501 atoms.
The the multi-solver scheme is constructed 
from the variational Wannier-state method for $\rho_A$
and the perturbative Wannier-state method $\rho_B$. \cite{THESIS}
An external load is imposed in the [001] direction 
and one initial defect bond is introduced 
by imposing a repulsive force on an atom pair. 
The sample is deformed with external load, 
the initial defect bond and thermal motion
but not fractured.  
The subsystems, 
$\rho_A$ and $\rho_B$, 
are assigned automatically during the MD simulation, 
as explained below; 
First, all the wavefunctions are calculated
by the perturbative solver method,
in which the weight of 
the unperturbed term $w_0^{(j)}$ is defined 
for each wavefunction $\phi_j$. (See Sec.~\ref{SEC-WANI-ON}) 
If the weight $w_0^{(j)}$ of a specific wavefunction
is less than 95 \% of the averaged weight  $w_0^{({\rm ave})}$
($w_0^{(j)} < 0.95 \, w_0^{({\rm ave})}$), 
the corresponding wavefunction $\phi_j$ 
is assigned into the subsystem $\rho_A$
and is determined by the variational procedure. 
In other words, if the perturbative procedure does not give 
a satisfactory accuracy, 
the procedure is switched automatically into the variational one.
The result of the automatic assignment is shown 
in Fig.~\ref{fig-4501-0716a-hyb-demo},
in which atoms are visible, only if 
its electron population is significantly contributed 
from $\rho_A$.
As a result,
the subsystem $\rho_A$, 
treated by the variational procedure, 
appear mainly near the sample edges and 
in the internal region near the initial defect bond,
because 
these regions are significantly deformed 
and the electronic states in these regions 
are fairly deviated from that in ideal crystal. 

As a technical detail 
of the MD simulation with the multi-solver scheme, 
we used a fine tuning technique of lattice constant; \cite{THESIS}
In calculations of ideal silicon crystal, 
the equilibrium lattice constant
or bond length differs by 2 \%  
between the variational and perturbative methods.
The difference can cause, in principle, 
an artificial lattice mismatch
in the multi-solver scheme
and therefore 
we tuned the bond length, 
by imposing 
an additional two-body classical potential 
on an atom pair or bond site, if the atom pair is occupied 
by a perturbative Wannier state.
This fine tuning technique avoids 
the possible artificial lattice mismatch. 
Although 
the calculation results without the fine tuning 
(not shown) 
did not indicate any practical problem
among our calculations of silicon,
we presume that
an error of 2 \% in lattice constant 
might be non-negligible in several cases. 
For example, the lattice constant 
between Si and Ge is different by 4 \% 
and the artificial lattice mismatch by 2 \% 
might cause a problem, 
when a Si/Ge system is calculated.

%\end{document}

\section{Applications}
\label{SEC-APPL}

%\begin{multicols}{2}
%-%-%-%-%-%-%-%-%-%-%-%-%-%-%-%-%-%-%-%-%-%-%-%-%-%-%-%-%-%-%
\begin{figure}[tbh]
\begin{center}
  \includegraphics[width=14cm]{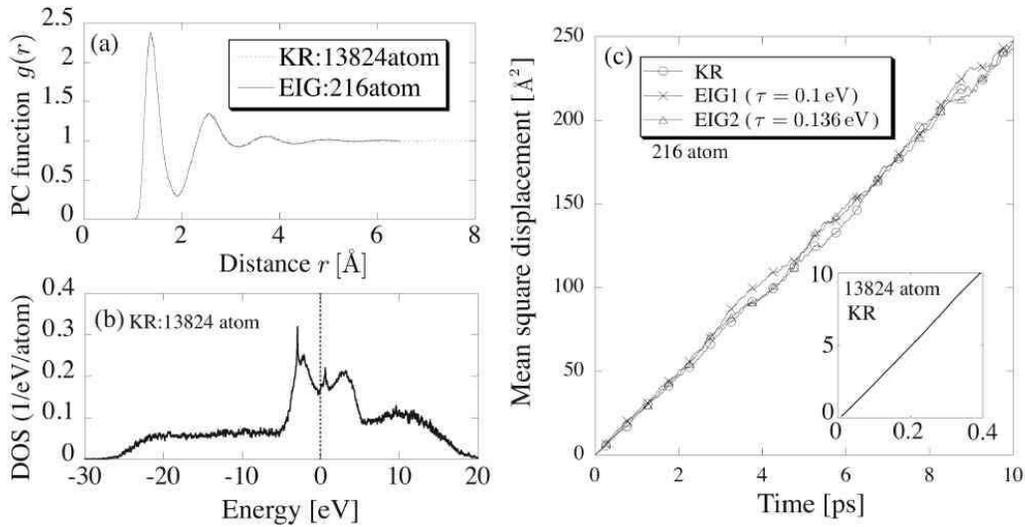}
\end{center}
\caption{
MD simulation of liquid carbon;
(a) Pair correlation (PC) function
calculated by the standard eigen-state method
with 216 atoms
and by the Krylov-subspace method 
with 13824 atoms.
In the former calculation,
the function is plotted only within  $r \le 6.45$\AA,
since the simulation cell size is smaller.
(b) DOS in a snapshot with 13824 atoms,
using the Krylov subspace method. 
The energy origin ($\varepsilon = 0$)
is chosen at the chemical potential.
(c) Mean square displacement 
using the Krylov subspace method (KR) or 
the standard eigen-state method.
In the latter method, the level-broadening 
parameter in the Fermi-Dirac function is set to 
$\tau=0.1$eV (EIG1) and $\tau=0.136$eV (EIG2). 
The numbers of atoms are 216 in the main figure
and 13824 in the inset, respectively.
}
\label{FIG-LIQ-C-RGR}
\end{figure}%
%-%-%-%-%-%-%-%-%-%-%-%-%-%-%-%-%-%-%-%-%-%-%-%-%-%-%-%-%-%-%

%%%%%%%%%%%%%%%%%%%%%%%%%%%%%%%%%%%%%%%%%
\begin{figure*}[thb]
\begin{center}
 \includegraphics[width=16cm]{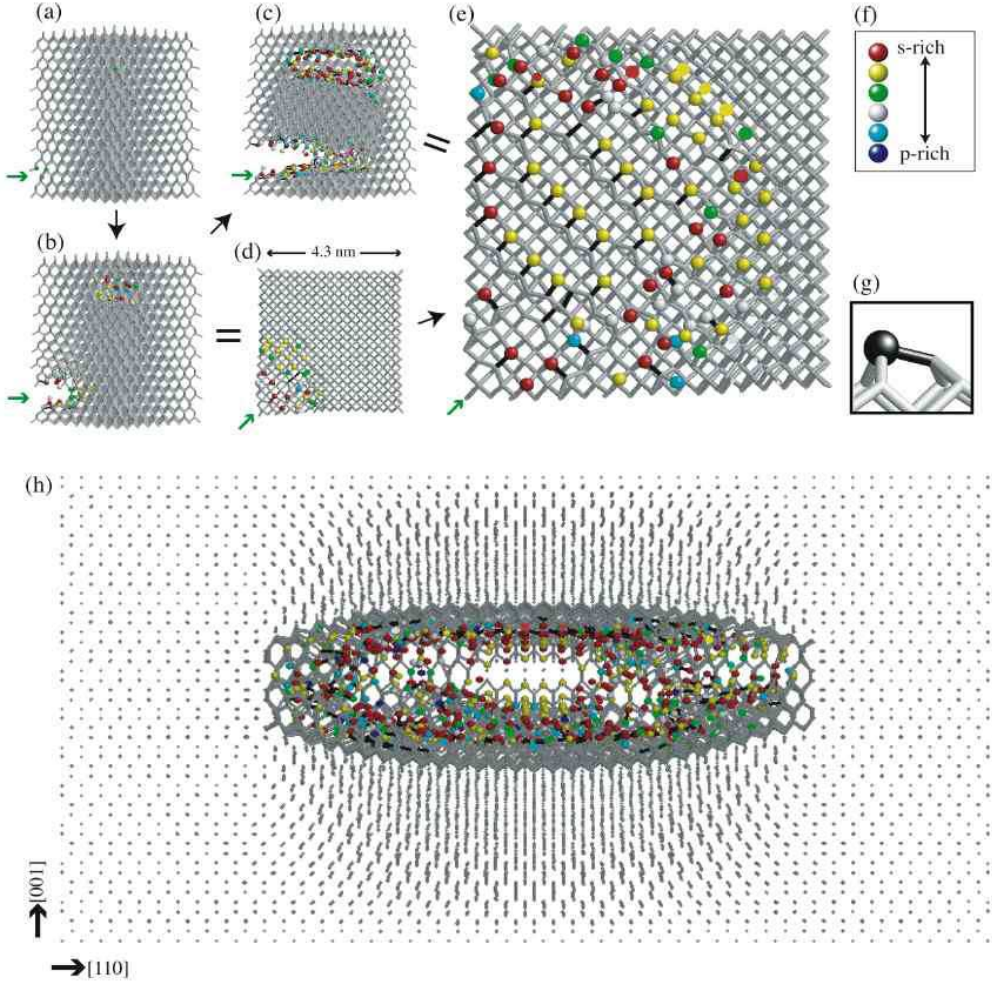}
\end{center}
\caption{
Cleavage process of a silicon nanocrystal under [001] external load,
in which a rod indicates a bonding state and  
a ball indicates an atomic (non-bonding) state.
(a)-(c):The 3D views of successive snapshots 
with the time interval of about 0.7 ps.
(d)(e):Top views of the lower cleavage plane, a (001) surface, 
in the snapshot (b) and (c), respectively.
The green arrow indicates the cleavage propagation direction
of the lower cleavage plane.
(f) Color samples of the atomic states (balls),
which indicate 
the weight of s orbitals ($f_{\rm s}^{(i)}$);
(i)$0 \le f_{\rm s}^{(i)} \le 0.2$ for blue, 
(ii)$0.2 \le f_{\rm s}^{(i)} \le 0.3$ for cyan, 
(iii)$0.3 \le f_{\rm s}^{(i)} \le 0.4$ for white, 
(iv)$0.4 \le f_{\rm s}^{(i)} \le 0.5$ for green, 
(v)$0.5 \le f_{\rm s}^{(i)} \le 0.6$ for yellow 
and (vi)$0.6 \le f_{\rm s}^{(i)} $ for red. 
(g) Example of the asymmetric dimer geometry.
(h)  A cleaved sample 
with 118850 atoms that is calculated 
by the multi-solver scheme. 
The picture is drawn 
for the  semi-infinite region of $y>x$. 
Electronic states, rods or balls, 
are depicted only for the  subsystem $\rho_A$.
Note that, in larger samples such as (h), 
the (001) cleavage mode will be fairly unstable, 
owing to step formations. \cite{HOSHI2003A}
}
\label{FIG-4501B}
\end{figure*}
%%%%%%%%%%%%%%%%%%%%%%%%%%%%%%%%%%%%%%%%%
% Note89,p.71,
%  NOTE89-1224a-2, nc011,20,29

\subsection{Liquid carbon : a metallic system}

Liquid carbon was simulated with 
the Krylov-subspace method as a test calculation.
The cubic simulation cell is used with 216 and 13824 atoms.
The density and the temperature are set to be 
$\rho = 2.0$ g/cm$^2$ and $T=6000$K, respectively. 
The time interval between MD steps is set to be $\Delta t =1$fs. 
As technical details, 
%In the Krylov subspace method, 
the subspace dimension  
and the number of atoms in the real-space projection 
(See \ref{SEC-APPENDIX-TECHNICAL})
are chosen to be $\nu=30$ and $N_{\rm PR}=200$, respectively.

Figure~\ref{FIG-LIQ-C-RGR}(a) shows 
the resultant pair correlation (PC) function
for the conventional eigen-state method
with 216 atoms
and for the Krylov-subspace method 
with 13824 atoms.
The two graphs are indistinguishable,
owing to an excellent agreement. 
In Fig.~\ref{FIG-LIQ-C-RGR}(b), 
the DOS is calculated, 
from the Green's function,
by the Krylov subspace method with 13824 atoms.
In the DOS calculation,
the controlling parameters are set 
into a heavier computational cost
($\nu=300$ and $N_{\rm RP}=1000$),
so as to reproduce the fine DOS profile.
Since the present Hamiltonian includes only $s$ and $p$ orbitals,
the resultant DOS is missing in higher energy regions.
The imaginary part of the energy $(z=E+i \gamma)$
is chosen at $\gamma =0.05$eV.
The resultant DOS profile in Fig.~\ref{FIG-LIQ-C-RGR}(b)
shows the correct feature of liquid carbon, as follows;
A narrow $\pi$ band appears, from $E = -5$eV to $+5$eV,
as in nanotubes,
which can be decomposed  
the bonding and antibonding bands. 
The $\pi$ bond in liquid phase is, however, 
imperfect and
non-bonding (atomic) $p$ states
appear as a sharp peak near the chemical potential
($\varepsilon \approx 0.6$eV).

Figure~\ref{FIG-LIQ-C-RGR}(c) shows 
the resultant mean square displacement (MSD)
for the Krylov subspace method (KR) and
the conventional  eigen-state method (EIG1,EIG2). 
The main figure  shows 
a system of 216 atoms by the two methods,
while the inset shows that of 13824 atoms
by the Krylov subspace method.
In the eigen-state method, 
the level-broadening (temperature) parameter 
in the Fermi-Dirac function is set to  
$\tau=0.1$eV (EIG1) and $\tau=0.005{\rm au}=0.136$eV (EIG2), 
respectively,
so as to show that 
the detailed treatment near the Fermi level 
causes different fluctuation behaviors of the MSD.
Since the difference in fluctuation behavior
is seen even among the two cases of the eigen-state method,
we conclude that 
the Krylov subspace method shows
satisfactory agreements with the eigen-state method
for PC function and diffusion constant 
(the gradient of the linear behavior 
in the main figure of Fig.~\ref{FIG-LIQ-C-RGR}(c)).

\subsection{Silicon :  cleavage process}
\label{APPL-SILICON-FRAC}

 As a practical large-scale calculation,
silicon cleavage process was investigated. 
\cite{HOSHI2003A,THESIS,HOSHI2005A}
The Wannier-state method is used, 
since it is faster than the Krylov-subspace method,
when, 
as discussed in Sec. \ref{SEC-COMPARI}, 
a dominant number of wavefunctions are well localized.
The number of atoms in the localization region 
for each Wannier state ($N_{\rm A}^{(i)}$)
is assigned to be $N_{\rm A}^{(i)}=20-80$,
which is determined by the residual norm $| \delta \phi_i^{\rm (WS)} |^2$.
%In result, the computational cost can be different among Wannier states. 
The resultant density matrix 
has a spatial spread, in its off-site elements,
over regions with hundreds of atoms.
Particularly,  
wavefunctions near cleaved regions tend to have a large residual norm 
and the localization constraint on such wavefunctions 
are automatically relaxed 
to increase the number $N_{\rm A}^{(i)}$. 
We found that 
such a way of controlling the accuracy for microscopic freedoms
is crucial for reproducing the surface reconstruction on cleaved surface. 
See Ref.~\cite{THESIS} for details.

Figure \ref{FIG-4501B}(a)-(c) 
shows a silicon cleavage process with the variational Wannier-state method.
The external load is imposed on the $[001]$ direction,
as in our previous simulation \cite{HOSHI2003A}.
The present system, unlike the previous one  \cite{HOSHI2003A},
does not contain any initial defect 
for \lq cleavage seed'.
In result, 
the cleavage starts from two points on the sample edges
and two cleavage planes appear. 
The lower cleavage surface is shown in Figs.~\ref{FIG-4501B}(d) and (e).
In Fig.~\ref{FIG-4501B},
a rod (atomic wavefunction) or ball (bonding wavefunction) 
is assigned 
for each wavefunction,
according to the weight distribution among atoms.
The black rods are the reconstructed bonds
that are not seen in the initial (crystalline) structure.
A ball is assigned for 
an atomic (non-bonding) orbital, localized on an atom site.
On the cleaved surface, 
an asymmetric dimer appears, as on a clean (001)  surface, 
with a ball (lone-pair state) on the upper atom, 
which is shown in Fig.~\ref{FIG-4501B}(g). 
For quantitative discussion of orbital freedoms,
a parameter $f_{\rm s}^{(i)}$ is defined, \cite{HOSHI2003A}
for a wave function $\phi_i$, as
\begin{eqnarray}
 f_{\rm s}^{(i)} \equiv   
 \sum_{I} | \langle \phi_i | I s \rangle |^2
 \label{FSDEF},
\end{eqnarray}
where $| I s \rangle $ is the s orbital at the $I$-th atom.
For example, $f_{\rm s}^{(i)} = 1/4$ in an ideal $sp^3$ hybridized state.
To visualize the orbital freedom of wave functions, 
the atomic (non-bonding) states are classified by the color of ball,
according to the value of $f_{\rm s}^{(i)}$
(See the caption of Fig.~\ref{FIG-4501B}). 
After a bulk (sp$^3$) bond is broken, 
the corresponding wavefunction is stabilized 
with increasing the weight of $s$ orbitals ($f_{\rm s}^{(i)} \ge 0.5$),
which results in appearance 
of red or yellow balls on cleaved surface.

As a remarkable result,  
a well-defined dimer-row domain 
is formed by nine dimers 
in Fig.~\ref{FIG-4501B}(e),
in which the tilting freedoms of asymmetric dimers are fixed
into the $(2 \times 1)$ configuration, 
although 
the surface energy of the $(2 \times 1)$ surface 
is higher than that of the $(4 \times 2)$ surface 
(See Sec. \ref{SEC-COMPARI}).  
We suggest that
the directional anisotropy of deformation is caused by 
the cleavage propagation direction, 
as indicated by the green arrow in Fig.~\ref{FIG-4501B}(e), 
and gives the ordering of the tilting freedoms 
into the $(2 \times 1)$ configuration. 
We also calculated many other systems (not shown)
in different sample geometry,
which supports the above suggestion.

Figure ~\ref{FIG-4501B}(h)  is a larger system simulated 
by the multi-solver scheme,
in which we use the variational 
and perturbative Wannier-state methods
for subsystems $\rho_A$ and $\rho_B$, respectively. 
\cite{HOSHI2003A,THESIS}
The system contains 118850 atoms and 
the sample dimension is 
$n_{110} \! \times \! n_{1\bar{1}0} \! \times \! n_{001} \! =
97 \times 100 \times 49 $
in the unit of the atomic layers,
where $n_{110}=100$ corresponds to about 20 nm.
Here the subsystem $\rho_A$ was composed of  
selected Wannier states near fracture regions
and the rest part of the electron system
is defined as the subsystem $\rho_B$. 
The number of Wannier states in 
the subsystem $\rho_A$ 
is approximately 5 \% of the total
and the computational cost 
by the present multi-solver scheme
is nearly $1/10$ of that 
%the same as that of $10^4$ atoms  
by the single-solver calculation with the variational procedure.
In Fig.~\ref{FIG-4501B}(h),
the electronic states in the subsystem $\rho_A$ 
are depicted as rods or balls
and those in the subsystem $\rho_B$ are invisible. 
The cleave surface in  Fig.~\ref{FIG-4501B}(h) contains 
(001) surface but is fairly unstable
with many step formations.~\cite{HOSHI2003A,THESIS}
See Refs.~\cite{HOSHI2003A,HOSHI2005A} 
for the physical discussion of the instability.

%%%%%%%%%%%%%%%%%%%%%%%%%%%%%%%%%%%%%%%%%
\subsection{General discussion on the multi-solver scheme}
\label{HYB-DISCUSS}

Finally, 
a general discussion is made
for a practical application of the multi-solver scheme. 
Among the present examples,
the procedure was carried out
without a self-consistent loop
($\rho_B \Rightarrow H_{\rm map}^{(A)} \Rightarrow \rho_A$), 
as explained in the beginning of 
Sec.~\ref{HYB-EX1}. 
The present non-selfconsistent procedure is practical, 
particularly, if  
the electron system can be decomposed into 
two parts that are governed 
by stronger and weaker binding mechanisms, respectively. 
In the present examples, 
the electronic states in the bulk part ($\rho_B$)
are governed by a stronger binding mechanism 
(the $sp^3$ bonding)
than those in surface states ($\rho_A$) and 
can be well described without any detailed 
information of the surface states ($\rho_A$).
Another example of the decomposition may be a system 
with strong $\sigma$ bonds and weak $\pi$ bonds. 
The situation of the decomposition 
is a candidate of the multi-solver scheme. 
When the multi-solver scheme is used,
the solver method for each subsystem 
should be chosen from the discussion 
of Sec.~\ref{SEC-COMPARI}.

We note that the multi-solver scheme with a self-consistent loop
can be realized, in principle,  and 
its practical application
might be a possible future work.

\section{Concluding remarks}

This paper presents 
fundamental theories and practical methods 
for large-scale electronic structure calculations,
particularly for 
dynamical process with nm-scale or 10-nm-scale structures.
First, we presented  
several practical solver procedures,  
based on Krylov subspace and generalized Wannier state, 
so as to obtain the density matrix without 
calculating eigen states.
We emphasized that 
every method has a way of accuracy control
for microscopic freedoms,
by monitoring the residuals of exact equations.
Second,   
the \lq multi-solver' scheme was formulated
based on the commutation relation of the density matrix
and was used for 
a hybrid or combined method of different solver methods.
Several practical large-scale calculations were carried out
in metallic and insulating cases.

These methodologies
enable us to design 
a simulation of nanostructure process
with an optimal computational cost, 
in which 
the accuracy  is controlled dynamically 
for microscopic (basis) freedoms and 
solver methods 
may be different among different regions. 
These points are essential in nanostructured systems,
nm-scale or 10-nm-scale systems, 
because 
a competition between different regions,
such as bulk and surface regions,
is essential and is required to be reproduced in simulation.
Since the above requirement is general among nanostructure processes,  
the present discussion is always valid, 
even when a different system 
is calculated by a different solver method 
from those in the present paper. 

\ack

This work is supported by a Grant-in-Aid from
the Ministry of Education, Science, Sports and Culture of Japan.
Numerical calculation was partly carried out 
using the facilities of the Japan Atomic Energy Research Institute, 
the Institute for Solid State Physics, University of Tokyo,
and the Research Center for Computational Science, Okazaki.

%%%%%%%%%%%%%%%%%%%%%%%%%%%%%%%%%%%%%%%%%%%%%%%%%%%%%%%%%%%
\begin{figure}[t]%hbp] 
\begin{center}
 \includegraphics[width=14cm]{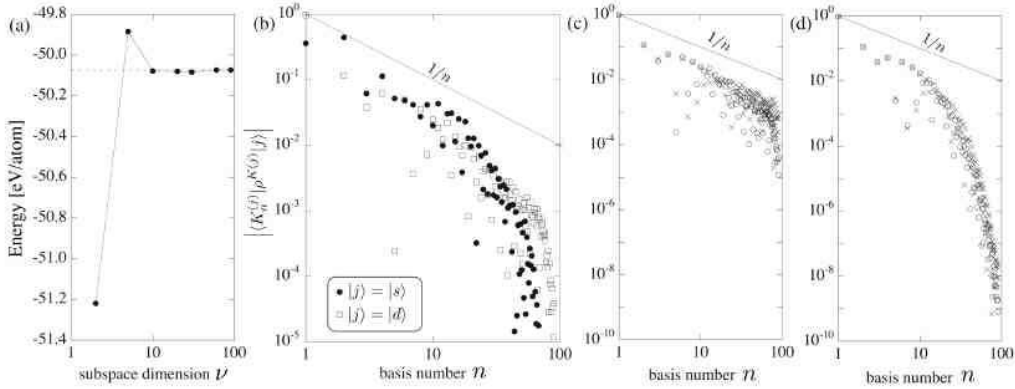}
\end{center}
\caption{\label{FIG-CU-DECAY} 
Krylov-subspace method for 
a fcc Cu system with 10,800 atoms; 
(a) The convergence behavior of the band structure energy 
as the function of the subspace dimension $\nu$ 
($\nu=$2, 5, 10, 20, 30, 60  and 90). 
A  reference value calculated by the standard eigen-state method 
is plotted as a dashed line.
(b) The decay behavior of 
the density matrix $\langle K_{n}^{(j)} |{\rho}_{}^{K(j)} |  j \rangle$,
as the function of the basis number ($n$).
The subspace dimension is set to be $\nu=90$
and the temperature (level-broadening) parameter is set to be 
$\tau=0.1$ eV.
The circle and square indicate
the values with the staring bases 
of the $s$ and $d$ ($e_g$) orbitals
($| j \rangle = | s \rangle, | e_g \rangle$), respectively.
(c)(d)Decay behavior 
of the density matrix 
$\langle K_{n}^{(j)} |{\rho}_{}^{K(j)} |  j \rangle$
with respect to the basis number 
of the Krylov subspace ($n$).
The circles 
indicate the result of 
10,800-atom system with the real-space projection
and the crosses 
indicate the result of 876-atom system 
without the real-space projection. 
The staring basis $( | j \rangle )$ is chosen to be 
a $d$ ($e_g$) orbital.
The temperature (level-broadening) parameter is set to be 
$\tau=0.1$ eV in (c) and 
$\tau=0.5$ eV in (d). 
}
\end{figure}
%%%%%%%%%%%%%%%%%%%%%%%%%%%%%%%%%%%%%%%%%%%%%%%%%%%%%%%%%%%

\appendix

%%%%%%%%%%%%%%%%%%%%%%%%%%%%%%%%%%%%%%%%%%%%%%%%%%%
\section{Proof of the fundamental equation in the multi-solver scheme}
\label{APPENDIX-HYB-WANI}

Here we prove Eq.~(\ref{HYB-EQN-A}),
the fundamental equation in the multi-solver scheme,
when the subsystems $\rho_A$, $\rho_B$
are constructed from Wannier states
in  Eqs.~(\ref{HYB-RHOA-EIG-DEV}) and (\ref{HYB-RHOB-EIG-DEV})
and the mapped Hamiltonian $H_{\rm map}^{(A)}$ is defined by 
Eq.~(\ref{HYB-HAM-1-ALL}).
We notice that the projection operator 
onto the unoccupied Hilbert space,
$\bar{\rho}$, is defined as
\begin{eqnarray}
 & & \bar{\rho} \equiv  1 - \rho 
 = 1 - \rho_A  - \rho_B
  \label{QMDM-UNOCC0} 
\end{eqnarray}
and satisfies
\begin{eqnarray}
 & & H \bar{\rho} = \bar{\rho} H.
  \label{QMDM-UNOCC1}
\end{eqnarray}
Equation ~(\ref{HYB-EQN-A}) is satisfied as follows; 
\begin{eqnarray}
 \left[ H_{\rm map}^{\rm (A)}, \, \, \rho_{\rm A} \right] 
 &=&  \left[ H, \rho_{\rm A} \right] 
  + 2 \eta_{\rm s} \left[ \rho_{\rm B}, \rho_{\rm A} \right] 
  - \left[H \rho_{\rm B}  +  
    \rho_{\rm B} H, \, \, \rho_{\rm A} \right] \no \\
 &=&	
	 \left( H \rho_{\rm A}  -  \rho_{\rm A} H \right) +0 
	-  \left( \rho_{\rm B} H \rho_{\rm A} - \rho_{\rm A} H \rho_{\rm B}
	    \right) \no \\
 &=& (1- \rho_{\rm B}) H  \rho_{\rm A}
        -  \rho_{\rm A} H \left( 1  - \rho_{\rm B} \right) \no \\
 &=& (\bar{\rho} + \rho_{\rm A}) H  \rho_{\rm A}
        -  \rho_{\rm A} H \left( \bar{\rho} + \rho_{\rm A} \right) \no \\
 &=& \bar{\rho} H  \rho_{\rm A} 
        - \rho_{\rm A} H \bar{\rho} \no \\
 &=& H \bar{\rho}  \rho_{\rm A} 
        - \rho_{\rm A} \bar{\rho} H =0,
 \label{QMDM-CALC-TOCHU}
\end{eqnarray}
where 
the second equality is obtained by 
Eqs.~(\ref{HYB-ORTHO}) and
the fourth and sixth equality is obtained by 
Eq.~(\ref{QMDM-UNOCC0}) and 
Eq.~(\ref{QMDM-UNOCC1}), respectively.
The last equality is obtained by
the orthogonality relation of $\bar{\rho} \rho_{\rm A}= \rho_{\rm A} \bar{\rho}=0$.

\section{Technical details and numerical aspects of the Krylov-subspace method}
\label{SEC-APPENDIX-TECHNICAL}

Here we discuss several technical details of the Krylov-subspace method
and demonstrate how the method works, particularly in metals. 
As a demonstration, 
a fcc Cu system was calculated
with a periodic simulation cell of $10,800$ atoms.
The temperature (level-broadening) parameter 
in the Fermi-Dirac function
is set to be $\tau=0.1$eV.
As a practical technique,
a real-space projection technique is introduced;
The Krylov subspace is generated 
by a Hamiltonian projected in real space, 
$H^{(j)} \equiv P^{(j)} H P^{(j)}$, 
instead of the original one $H$,
where the projection operator $P^{(j)}$ 
projects a function onto 
the spherical region whose center is located 
at the atomic position of the $j$-th atomic basis. 
The resultant Krylov subspace is the same as the original one
$({\cal K}_n(H, | j \rangle) = {\cal K}_{n}(H^{(j)}, | j \rangle)$, 
while the bases lie within the projection region
( $(H^{(j)})^n | j \rangle =  H^n | j \rangle$). 
Since the procedure of constructing the Krylov subspace 
${\cal K}_{\nu}(H^{(j)}, | j \rangle)$
is independent among the starting bases $(j)$,
all the procedures and the quantities are well-defined 
with the real-space projection technique.
The projection radius is  determined 
for each starting basis $| j \rangle$,
so that 
a given number of atoms, $N_{\rm RP}$, 
should be contained inside the radius. 
The present calculation with $10,800$ atoms was carried out 
using the projection technique with $N_{\rm RP}=381$.
The calculation without the projection technique 
was also carried out in a smaller (876-atom) system,
which is discussed below.

In Fig.~\ref{FIG-CU-DECAY}(a),
the convergence behavior 
of the calculated band structure energy is shown 
as the function of the subspace dimension, 
in which a  reference value is also calculated 
by standard eigen-state calculation 
with the standard Brouillion-zone integration.
The deviation from the reference value 
is about 0.01 eV per atom for $\nu=$10, 20 and 30 
and less than 1 meV per atom for $\nu=60$ and 90.
Since the density matrix $\rho_{ij}$ is calculated 
in the form of Eq.~(\ref{EQ-DM-SUMMATION}),
its representation  within the Krylov subspace
$\langle K_{n}^{(j)} |{\rho}_{}^{K(j)} |  j \rangle 
(= \langle K_{n}^{(j)} |{\rho}_{}^{K(j)} |  K_{1}^{(j)} \rangle)$ 
is plotted in Fig.~\ref{FIG-CU-DECAY}(b),
where the starting bases $| j \rangle $ are set to be
$s$ and $d$ ($e_{g}$) orbitals, as examples.
In Fig.~\ref{FIG-CU-DECAY}(b), we observe a $1/n$ or faster decay, 
and this observation is also seen
with the other staring bases ($p$ and $t_{2g}$ orbitals). 
The decay behavior of Fig.~\ref{FIG-CU-DECAY}(b) 
is explained by a general mathematical analysis of the Lanczos procedure 
~\cite{TAKAYAMA2004A}, 
in which a $1/n$ decay should appear 
with the zero-temperature formulation ($\tau=0$) 
and a faster decay should appear 
with a finite temperature formulation ($\tau \ne 0$).
The quantity $ \langle i |  K_{n}^{(j)} \rangle$, on the other hand,
also decays as $1/n$ or faster (not shown),
since the (normalized) vector $| K_{n}^{(j)} \rangle$
has a spatial spread within $n$-th hopping range from the starting basis 
($| K_n^{(j)} \rangle \in {\cal K}_{n}(H, | j \rangle)$). 
Consequently, their product 
($\langle i |  K_{n}^{(j)} \rangle \langle K_{n}^{(j)} |{\rho}_{}^{K(j)} |  j \rangle$) 
decays as $1/n^2$ or faster,
which validates the fast convergence in the summation 
of Eq.~(\ref{EQ-DM-SUMMATION}).

We should emphasis that
the decay behavior in Fig.~\ref{FIG-CU-DECAY}(b)
comes from a general property of the Lanczos procedure,
as discussed above, 
{\it not} from the projection technique. 
The above statement is confirmed numerically in 
Figs.~\ref{FIG-CU-DECAY}(c)(d),
in which  fcc Cu systems
were calculated with or without 
the real-space projection
and  the resultant decay behavior 
is affected significantly by the temperature 
(level-broadening) parameter $\tau$,
but not by the  projection technique.

%%%%%%%%%%%%%%%%%%%%%%%%%%%%%%%%%%%%%%%%%

\newcommand{\noop}[1]{}

\end{document}